\documentstyle[a4]{article}
\def\abstract#1{\vskip 7mm 
        \begin{center}{\large Abstract}\par \smallskip
                \begin{minipage}[c]{12cm}
                        \small #1
                \end{minipage}
        \end{center}
}
\def\title#1{\begin{center}{\Large\bf #1}\end{center}}
\def\author#1{\vskip 5mm \begin{center}{#1}\end{center}}
\def\address#1{\begin{center}{\it #1}\end{center}}
\begin{document}

\def\R{{\bf R}}
\def\Z{{\bf Z}}
\def\del{\partial}
\def\Lap{\bigtriangleup}
\def\Div{{\rm div}\ }
\def\rot{{\rm rot}\ }
\def\curl{{\rm curl}\ }
\def\grad{{\rm grad}\ }
\def\Tr{{\rm Tr}}
\def\^{\wedge}
\def\goinf{\rightarrow\infty}
\def\goes{\rightarrow}
\def\bm{\boldmath}
\def\-{{-1}}
\def\inv{^{-1}}
\def\sqr{^{1/2}}
\def\isqr{^{-1/2}}
\def\wa{\!\!\!\!&=&\!\!\!\!}
\def\wb{\!\!\!\!&\equiv &\!\!\!\!}

\def\reff#1{(\ref{#1})}
\def\vb#1{{\partial \over \partial #1}} % vector basis
\def\Del#1#2{{\partial #1 \over \partial #2}}
\def\Dell#1#2{{\partial^2 #1 \over \partial {#2}^2}}
\def\Dif#1#2{{d #1 \over d #2}}
\def\Lie#1{ {\cal L}_{#1} }
\def\diag#1{{\rm diag}(#1)}
\def\abs#1{\left | #1 \right |}
\def\rcp#1{{1\over #1}}
\def\paren#1{\left( #1 \right)}
\def\brace#1{\left\{ #1 \right\}}
\def\bra#1{\left[ #1 \right]}
\def\angl#1{\left\langle #1 \right\rangle}
\def\matrix#1#2#3#4#5#6#7#8#9{
        \left( \begin{array}{ccc}
                        #1 & #2 & #3 \\ #4 & #5 & #6 \\ #7 & #8 & #9
        \end{array}     \right) }
\def\manbo#1{}  % For comments.

\def\d{{\rm d}}
\def\c#1{{}_{,#1}}
\def\tta#1#2{\theta^{#1}{}_{#2}}
\def\itta#1#2{\theta_{#1}{}^{#2}}

\def\hh{{h}}
\def\g{{g}}
\def\uh#1#2{\hh^{#1#2}}
\def\dh#1#2{\hh_{#1#2}}
\def\ug#1#2{\g^{#1#2}}
\def\dg#1#2{\g_{#1#2}}

\title{
On the null surface formalism \\
\smallskip 
{\large -- Formulation in three dimensions and gauge freedom --}
}
\author{
        Masayuki Tanimoto\footnote{E-mail:tanimoto@yukawa.kyoto-u.ac.jp}
}
\address{
        Yukawa Institute for Theoretical Physics,  
        Kyoto University, \\ 
        Kyoto 606-01, Japan
}
\abstract{
The null surface formalism of GR in three dimensions is presented,
and the gauge freedom thereof, which is not just diffeomorphism, is
discussed briefly.
}

%%%%%%%%%%%%%%%%%%%%%%%%%%%%%%%%%%%%%%%
\section{Introduction}
%%%%%%%%%%%%%%%%%%%%%%%%%%%%%%%%%%%%%%%

Recently, Frittelli, Kozameh, and Newman \cite{FKN1,FKN2,FKN3} presented
an unconventional version of GR, where the variable is not any field
such as the metric or connection, but 2-parameters family of foliations
\begin{equation}
        Z(x^a;\zeta,\bar\zeta)={\rm const.}
        \label{1-1}
\end{equation}
of a manifold, where $x^a$ is an arbitrary coordinate system, and
$\zeta$ and $\bar\zeta$ are parameters.  In this formalism, called the
{\it null surface formalism (NSF)}, part of main equations, called the
{\it metricity condition}, is obtained from the requirement that the
hypersurfaces $Z(x^a;\zeta,\bar\zeta)={\rm const.}$ be {\it null} with
respect to a metric.  This condition not only says the existence of a
corresponding metric on the manifold, but also contains information of
null geodesics.  (For an attempt toward quantization, see
Refs.\cite{q1,q2}.)

We shall in this article present the three-dimensional version of NSF.
An establishment of NSF in dimensions other than four was nontrivial at
the point the original formulation was presented, since the NSF is not a
conventional ``field'' theory.  Our success in three dimensions may
suggest that NSF can establish in any dimensions equal to or higher than
three.  Moreover, the simplicity in three dimensions makes the algebraic
structure of the NSF transparent.  We use our formalism to discuss
``gauge'' freedom of NSF, which is also rather unconventional.

%%%%%%%%%%%%%%%%%%%%%%%%%%%%%%%%%%%%%%%
\section{Null foliations and the intrinsic coordinates}
%%%%%%%%%%%%%%%%%%%%%%%%%%%%%%%%%%%%%%%

Consider a Lorentzian metric $\ug ab(x^a)$ on a three dimensional
manifold $M$, and 1-parameter family of null foliations of $M$;
\begin{equation}
        Z(x^a,\zeta)={\rm const.},
        \label{2-1}
\end{equation}
where $x^a$ is an arbitrary coordinate system, and $\zeta$ is a real
parameter.

As is emphasized in Refs.\cite{FKN1,FKN2}, a family of null foliations
defines a particular coordinate system $\theta^i$, called the {\it
  intrinsic coordinates}, which is given by successive derivatives of
$Z$ with respect to the parameter $\zeta$;
\begin{eqnarray}
        \theta^0\wb u\equiv Z(x^a,\zeta),       \nonumber \\
        \theta^1\wb \omega\equiv \del Z(x^a,\zeta),     \nonumber \\
        \theta^2\wb R\equiv \del^2 Z(x^a,\zeta),
         \label{2-2}
\end{eqnarray}
where
\begin{equation}
        \del\equiv{\del\over\del\zeta}.
        \label{2-3}
\end{equation}
We may write as
\begin{equation}
        \theta^i=\del^iZ(x^a,\zeta).
        \label{2-4}
\end{equation}
For later convenience, we define the ``Jacobian''
\begin{equation}
        \tta ia\equiv\theta^i\c a,\quad \tta ia\itta ja\equiv\delta^i_j.
        \label{2-4.1}
\end{equation}
We also define
\begin{equation}
        \Lambda\equiv\del^3Z(\theta^i,\zeta).
        \label{2-8}
\end{equation}

We can find the components of the metric with respect to $\theta^i$
\begin{equation}
        \ug ij=\ug ab \del^iZ\c a\del^jZ\c b,
        \label{2-5}
\end{equation}
by successively operating $\del$ on the null condition of $Z(x^a,\zeta)$
\begin{equation}
        \ug00=\ug ab(x^a)Z\c aZ\c b=0.
        \label{2-6}
\end{equation}
For example, the first derivative of Eq.\reff{2-6}, $\ug ab\del Z\c aZ\c
b=0$, immediately implies $\ug01=0$.  If we define
\begin{equation}
        \Omega^2\equiv\ug02,
        \label{omega}
\end{equation}
which is independent of $Z(x^a,\zeta)$, the final result is
\begin{equation}
        \ug ij=\Omega^2
        \matrix 0010{-1}{-(1/3)\Lambda\c2}1{-(1/3)\Lambda\c2}
        {-(1/3)\del(\Lambda\c2)+(1/9)(\Lambda\c2)^2+\Lambda\c1}.
        \label{2-7}
\end{equation}
Here, the operation of $\del$ on a function $f(\theta^i,\zeta)$ is
well-defined;
\begin{equation}
        \del f(\theta^i,\zeta)=\del'f+(\del\theta^i)f\c i
        =\del'f+\omega f\c0+R f\c1+\Lambda f\c2,
        \label{2-9}
\end{equation}
where $\del'$ is the differential operator for fixed $\theta^i$.

With Eq.\reff{2-7}, we can check that $l^a\equiv(\del/\del R)^a$ is a
null geodesic generator;
\begin{equation}
        l^al_a=0,\quad l^b\nabla_bl^a=-2\frac{\Omega\c2}{\Omega}l^a.
        \label{2-10}
\end{equation}

\section{The metricity conditions and Einstein's equation}

As in \cite{FKN2}, the metricity conditions can be expressed as
\begin{eqnarray}
        && \del\ug ab(x^a)=0 \nonumber \\
        &\Leftrightarrow & \del(\ug ij(\theta^i,\zeta)\itta ia\itta jb)=0
                \nonumber \\
        &\Leftrightarrow &
                \del\ug ij(\theta^i,\zeta)+
                \ug km(\theta^i,\zeta)\del(\itta ka\itta mb)\tta ia\tta jb=0.
        \label{3-1}
\end{eqnarray}
That is, we demand that the $\zeta$-dependence of $\ug
ij(\theta^i,\zeta)$ can be ``absorbed'' into a coordinate transformation
of the form \reff{2-2}.  It is important to note that $\theta^i$
themselves are regarded as $\zeta$-dependent, so $\ug
ij(\theta^i,\zeta)=\ug ij(\theta^i(\zeta),\zeta)$.  This implies that
even though $\ug ij$ has no explicit dependence of $\zeta$, the
metricity conditions are nontrivial.

\def\udT#1#2{T^{#1}{}_{#2}} To proceed, we define matrix $\udT ij$ by
\begin{equation}
        \del\tta ia\equiv\udT ij\tta ja.
        \label{3-2}
\end{equation}
With this, Eq.\reff{3-1} can be written as
\begin{equation}
        \del\ug ij-\udT ik\ug kj-\udT jk\ug ki=0.
        \label{3-3}
\end{equation}
We can easily have the explicit form of $\udT ij$.  For example, noting
$\tta ia=\theta^i{}_{,a}$, we have for $i=0$ for Eq.\reff{3-2}
\begin{eqnarray}
        && \del Z_{,a}=\udT 0j\theta^j{}_{,a} \nonumber \\
        &\Leftrightarrow& \theta^1{}_{,a}=\udT 0j\theta^j{}_{,a},
        \label{3-4}
\end{eqnarray}
so that
\begin{equation}
        \udT 0j=\delta^1_j.
        \label{3-5}
\end{equation}
Similarly, we have for other components
\begin{equation}
        \udT 1j=\delta^2_j,\quad \udT 2j=\Lambda_{,j}.
        \label{3-6}
\end{equation}

We have six conditions Eq.\reff{3-3} (under Eqs.\reff{3-5} and
\reff{3-6}), and moreover we impose the null condition $\ug 00=0$ and
define Eq.\reff{omega}.  We therefore have eight inputs, so eight
outputs must follow.  Six of them are given by the components of $\ug
ij$, Eq.\reff{2-7}.  The rest is the metricity conditions we searched
for, which is the following two consistency conditions;
\begin{equation}
        \Lambda\c2=3\Omega\inv\del\Omega,
        \label{M1}
\end{equation}
and
\begin{equation}
        -{4\over9}(\Lambda\c2)^3+2\Lambda\c2\del(\Lambda\c2)
        -2\Lambda\c1\Lambda\c2
        -\del^2(\Lambda\c2)+3\del(\Lambda\c1)-6\Lambda\c0=0.
        \label{M2}
\end{equation}

If we find a solution of Eqs.\reff{M1} and \reff{M2} for $\Lambda$ and
$\Omega$, we obtain a metric \reff{2-7}, and simultaneously, have
essentially all information of null geodesics thereof (cf.
Eq.\reff{2-10}).

Next, we consider Einstein's equation with cosmological constant
$\lambda$,
\begin{equation}
        G^{ab}=\lambda\ug ab,
        \label{4-1}
\end{equation}
where $G^{ab}$ is the Einstein tensor.

We cast the gradient of $Z$ to Eq.\reff{4-1};
\begin{eqnarray}
        && G^{ab}Z\c aZ\c b=\lambda\ug abZ\c aZ\c b=0 \nonumber \\
        &\Leftrightarrow& G^{00}=0  \nonumber \\
        &\Leftrightarrow& G_{22}=0.
        \label{4-2}
\end{eqnarray}
From a straightforward calculation, $G_{22}=\Omega\inv\Omega\c{22}$.  We
thus have
\begin{equation}
        \Omega\inv\Omega\c{22}=0.
        \label{E1}
\end{equation}

If we operate $\del$ on (the first row of) Eq.\reff{4-2}, we obtain
\begin{equation}
        G^{11}+G^{02}=0,
        \label{4-4}
\end{equation}
but this does {\it not} mean the two components of Einstein's equation
for $(ij)=(11)$ and $(02)$, given by
\begin{equation}
        G^{11}=-\lambda\Omega^2
        \label{4-5}
\end{equation}
and
\begin{equation}
        G^{02}=\lambda\Omega^2.
        \label{4-6}
\end{equation}
Therefore, we need impose one of the two equations independently of
Eq.\reff{4-2} (or Eq.\reff{E1}).  \footnote{Ref.\cite{FKN2} lacks the
  corresponding equation.} We take Eq.\reff{4-6}, which is, after some
calculations, found to be
\begin{eqnarray}
        &&
        -{1\over36}\Omega^2( (\Lambda\c{22})^2+6\Lambda\c{221} )
        +{1\over9}(\Omega\c2)^2( -3\del(\Lambda\c{2})
        +(\Lambda\c2)^2+9\Lambda\c1 )
        \nonumber \\
        &&
        +{1\over18}\Omega\Omega\c2( 3\del(\Lambda\c{22})
        +\Lambda\c2\Lambda\c{22} )
        +{2\over3}\Omega\c2( 3\Omega\c0-\Lambda\c2\Omega\c1 )
        \nonumber \\
        &&
        +{1\over6}\Omega( \Lambda\c{22}\Omega\c1+2\Lambda\c2\Omega\c{12}
        +6\Omega\c{11}-6\Omega\c{02} )
        -(\Omega\c1)^2
        =\lambda.
        \label{E2}
\end{eqnarray}
We have used, as well as Eq.\reff{E1}, identity
\begin{equation}
        (\del(\Lambda\c2))\c2=
        \del(\Lambda\c{22})+\Lambda\c{12}+\Lambda\c2\Lambda\c{22},
        \label{4-7}
\end{equation}
obtained from more general identity for an arbitrary function $\phi$
\begin{equation}
        \del(\phi\c2)-(\del\phi)\c2=-\phi\c1-\Lambda\c2\phi\c2.
        \label{4-8}
\end{equation}

We can check that any other components of Einstein's equation follow
from the successive differentiations with respect to $\zeta$ of
Eq.\reff{4-4}.  Thus, we need only two component equation, Eqs.\reff{E1}
and \reff{E2}, rather than six components.

Our fundamental equations are therefore the four coupled nonlinear
partial differential equations \reff{M1}, \reff{M2}, \reff{E1}, and
\reff{E2}.  The following fact may be noteworthy at this point.  First,
we can immediately solve Eq.\reff{E1}, and find that $\Omega$ is a
polynomial of variable $R$ at most of first order.  Then, it is also
easy to see from Eq.\reff{M1} that $\Lambda$ is a polynomial of variable
$R$ at most of third order.  Thus, both $\Lambda$ and $\Omega$ are
polynomials of $R$ in {\it any} ``gauge'', discussed in the next
section.

\section{``Gauge'' degrees of freedom}
\label{gaugesection}

For a fixed spacetime, we can consider many varieties of 1-parameter
family of null foliations, and which, in general, correspond to
different $\Lambda$ and $\Omega$. Therefore, even in the present
formalism there exists a sort of ``gauge'' freedom.

More precisely, we can categorize this ``gauge'' into two parts, one is
essential {\it deformations} of way of taking the 1-parameter family of
null foliations, and the other is just {\it reparametrizations} of the
1-parameter family of null foliations.

We, in this section, discuss two simple cases of reparametrizations.

\subsection{Reparametrization of $\zeta$}

Consider a 1-parameter family of foliations $u={\rm
  const.}=Z(x^a,\zeta)$.  Reparametrization of the form
\begin{equation}
        \zeta\goes f(\zeta),
        \label{5-1}
\end{equation}
where $f$ is a real function, apparently preserves the original
foliation.  Hence, this is a sort of gauge.

Under the transformation \reff{5-1}, we have
\begin{equation}
        Z(x^a,\zeta)\goes \tilde Z(x^a,\zeta)=Z(x^a,f(\zeta)).
        \label{5-2}
\end{equation}
We denote the new intrinsic coordinates and $\Lambda$ obtained from
$\tilde Z$ as $(\tilde u,\tilde\omega,\tilde R)$ and $\tilde\Lambda$,
respectively.  Operating $\del$ successively on $\tilde Z$, we can
easily have
\begin{eqnarray}
        && \tilde u=u,\quad \tilde\omega=\omega f'(\zeta), \nonumber \\
        && \tilde R=R f'(\zeta)^2+\omega f''(\zeta), \nonumber \\
        && \tilde\Lambda= 
                \Lambda f'(\zeta)^3+3R f'(\zeta)f''(\zeta)+\omega f'''(\zeta).
         \label{5-3}
\end{eqnarray}
Thus, we have obtained the transformation rule for $\Lambda$.  The rule
for $\Omega$ can also be obtained, if we think of Eq.\reff{5-3} as a
coordinate transformation $\theta^i\goes\tilde\theta^i$.

From the Jacobian
\begin{equation}
        \tilde\theta^i{}_{,k}=
        \matrix 1000{f'(\zeta)}00{f''(\zeta)}{f'(\zeta)^2},
        \label{5-4}
\end{equation}
we have
\begin{eqnarray}
        \tilde g^{ij}\wa\ug kl\tilde\theta^i{}_{,k}\tilde\theta^j{}_{,l}
        \nonumber \\
        \wa f'(\zeta)^2\Omega^2\matrix 0010{-1}{*}1{*}{**},
        \label{5-5}
\end{eqnarray}
where
\begin{equation}
        *\equiv -{f''(\zeta)\over f'(\zeta)}-\rcp3 f'(\zeta)\Lambda\c2
        \label{5-5a}
\end{equation}
and
\begin{equation}
        **\equiv -\paren{{f''(\zeta)\over f'(\zeta)}}^2
        -\frac23 f''(\zeta)\Lambda\c2
        +f'(\zeta)^2\paren{-\rcp3\del\Lambda\c2
          +\Lambda\c1+\rcp9(\Lambda\c2)^2}.
        \label{5-6}
\end{equation}
Comparing with Eq.\reff{2-7}, we find
\begin{equation}
        \Omega\goes\tilde\Omega=f'(\zeta)\Omega.
        \label{5-7}
\end{equation}
That is, $\Omega$ so transforms as $\omega$.  (We can check that $*$ and
$**$ are certainly of the form of Eq.\reff{2-7}, if written in terms of
the tilded variables.)

\subsection{Reparametrization of $u$}

Consider a foliation
\begin{equation}
        {\rm const.}=Z(x^a).
        \label{6-1}
\end{equation}
Then, the foliation
\begin{equation}
        {\rm const.}=\phi(Z(x^a)),
        \label{6-2}
\end{equation}
where $\phi$ is a real function, is geometrically the same foliation as
the original.

We thus consider the transformation
\begin{equation}
        Z(x^a,\zeta)\goes \tilde Z(x^a,\zeta)=\phi(Z(x^a,\zeta)).
        \label{6-3}
\end{equation}
The following procedure to obtain the transformation rule for $\Lambda$
and $\Omega$ is exactly the same as before.  We denote the new intrinsic
coordinates and $\Lambda$ obtained from $\tilde Z$ as $(\tilde
u,\tilde\omega,\tilde R)$ and $\tilde\Lambda$, respectively.  Operating
$\del$ successively on $\tilde Z$, we have
\begin{eqnarray}
        && \tilde u=\phi(u),\quad \tilde\omega=\phi'(u)\omega, \nonumber \\
        && \tilde R=\phi''(u)\omega^2+\phi'(u)R, \nonumber \\
        && \tilde\Lambda= 
                \phi'''(u)\omega^3+3\phi''(u)\omega R+\phi'(u)\Lambda.
         \label{6-4}
\end{eqnarray}
The rule for $\Omega$ can also be obtained by observing the coordinate
transformation $\theta^i\goes\tilde\theta^i$.  We find
\begin{equation}
        \Omega\goes\tilde\Omega=\phi'(u)\Omega.
        \label{6-5}
\end{equation}
Again, $\Omega$ transforms like $\omega$.

\section{An example}

As an example of solution, we consider the Minkowski space, which is the
only (local) vacuum solution of Einstein's equation in three dimensions.

The metric is, in the standard coordinates, given by
\begin{equation}
        \d s^2=\d t^2-\d x^2-\d y^2.
        \label{7-1}
\end{equation}
From the 1-parameter family of (or ``$S^1$'s worth'' of) null vector
fields given by \def\v#1{\paren{{\del\over\del #1}}}
\begin{equation}
        l^a=\v t^a-\cos\zeta\v x^a-\sin\zeta\v y^a,
        \label{7-2}
\end{equation}
we can easily have a 1-parameter family of null foliations;
\begin{equation}
        u=Z(x^a,\zeta)=l_ax^a=t+x\cos\zeta+y\sin\zeta.
        \label{7-3}
\end{equation}
Operating $\del^n$, we have
\begin{eqnarray}
        \omega\wa\del Z=-x\sin\zeta+y\cos\zeta, \nonumber \\
        R\wa\del^2 Z=-x\cos\zeta-y\sin\zeta, \nonumber \\
        \Lambda\wa\del^3 Z=x\sin\zeta-y\cos\zeta.
        \label{7-4}
\end{eqnarray}
We, thus, have
\begin{equation}
        \Lambda=-\omega.
        \label{7-5}
\end{equation}
As for $\Omega$, like in the previous section, thinking of
Eqs.\reff{7-3} and \reff{7-4} as a coordinate transformation, we have
\begin{equation}
        \Omega=1,
        \label{7-6}
\end{equation}
since the metric in terms of the intrinsic coordinates is found to be
\begin{equation}
        \d s^2=\d u^2+2\d u\d R-\d \omega^2.
        \label{7-7}
\end{equation}
We can easily check that Eqs.\reff{7-5} and \reff{7-6} satisfy the
fundamental equations \reff{M1}, \reff{M2}, \reff{E1}, and \reff{E2}.
Also, we can check $l^a=(\del/\del R)^a$.

\section{Conclusion}

We have formulated the three dimensional version of NSF.  It inherits
all the properties of the original one, but is, of course, much simpler.
We have explicitly discussed two simple classes of gauge with our simple
version of NSF.  As we have seen, gauge in NSFs is not just
diffeomorphism of the manifold --- it presumably corresponds to a
subclass of the diffeomorphisms, but is unclear up to now.

%%%%%%%%%%%%%%%%%%%%%%%%%%%%%%%%%%%%%%%
\section*{Acknowledgments}

I would like to thank Professor H. Kodama for helpful comments, T. Koike
for an enlightening conversation, and M. Siino for discussions at an
early stage of this work.
%%%%%%%%%%%%%%%%%%%%%%%%%%%%%%%%%%%%%%%

%%%%%%%%%%%%%%%%%%%%%%%%%%%%%%%%%%%%%%%

\end{document}